\documentclass[12pt,preprint]{aastex}

\shorttitle{Spontaneous Onset of Fast Magnetic Reconnection}
\shortauthors{Cassak, Drake, and Shay}

\begin{document}


\title{A Model for Spontaneous Onset of Fast Magnetic Reconnection}


\author{P.~A.~Cassak\altaffilmark{1}, J.~F.~Drake\altaffilmark{1}
and M.~A.~Shay\altaffilmark{2}}


\altaffiltext{1}{Institute for Research in 
Electronics and Applied Physics, 
University of Maryland, College Park, MD 20742;
pcassak@glue.umd.edu, drake@plasma.umd.edu}

\altaffiltext{2}{Department of Physics and Astronomy,
University of Delaware, Newark, DE, 19716;
shay@physics.udel.edu}


\begin{abstract}
We present a model for the spontaneous onset of fast magnetic
reconnection in a weakly collisional plasma, such as the solar corona.
When a current layer of macroscopic width undergoes collisional
(Sweet-Parker) reconnection, a narrow dissipation region forms around
the X-line.  This dissipation region naturally becomes narrower during
the reconnection process as stronger magnetic fields are convected
toward the X-line.  When the dissipation region becomes thinner than
the ion skin depth, resistive magnetohydrodynamics breaks down as the
Hall effect becomes important and the Sweet-Parker solution ceases to
exist.  A transition to collisionless (Hall) reconnection ensues,
increasing the reconnection rate by many orders of magnitude in a very
short time.  Predictions of the model are consistent with constraints
set by observations of solar flares.
\end{abstract}





\section{INTRODUCTION}

Magnetic reconnection is the driver of explosions in the solar corona.
The first self-consistent description of magnetic reconnection, the
Sweet-Parker model \citep{Sweet58,Parker57}, was immediately
recognized as yielding energy release rates too slow to explain
observations.  Petschek reconnection \citep{Petschek64} and related
models \citep{Priest86} can be much faster, but requires anomalous
resistivity \citep{Sato79,Biskamp86}, a process which is not well
understood.  In the past 10-15 years, a new paradigm of collisionless
(Hall) reconnection has emerged, in which non-magnetohydrodynamic
terms make reconnection much faster \citep{Birn01}, about six orders
of magnitude faster for solar flare parameters.  Signatures of Hall
reconnection have been observed in magnetospheric observations
\citep{Oieroset01,Mozer02} and laboratory experiments
\citep{Cothran03,Ren05}.

However, explaining observed reconnection rates is only part of the
problem.  One must also explain why reconnection is explosive:
explaining how magnetic energy can accumulate without significant
dissipation and identifying the mechanism triggering the onset of fast
(Hall) reconnection to release the stored energy are long-standing
problems.

Recently, \citet{Cassak05} showed that a catastrophic transition from
Sweet-Parker to Hall reconnection occurs when the width $\delta$ of
the Sweet-Parker dissipation region falls below the ion skin depth
$d_{i} = c / \omega_{pi}$, where $\omega_{pi} = (4 \pi n e^{2} /
m_{i})^{1/2}$ is the ion plasma frequency and $n$ is the plasma
density.  This is the length scale at which magnetohydrodynamics (MHD)
breaks down and the Hall term in the generalized Ohm's law becomes
important.  For $\delta < d_{i}$, the Sweet-Parker solution ceases to
exist, and an abrupt transition to Hall reconnection ensues.

The catastrophic transition to Hall reconnection when $\delta \sim
d_{i}$ was demonstrated by (externally) decreasing the resistivity
$\eta$.  However, the idea that a solar eruption is caused by a change
in the resistivity by a large enough amount to cause a transition from
collisional to collisionless reconnection requires ad hoc assumptions
about the poorly understood energetics of the corona.  In this letter,
we suggest that the transition can occur as the result of the
dissipation region thinning due to the convection of stronger magnetic
fields into the dissipation region during slow Sweet-Parker
reconnection.  This is a generic process which is entirely
self-driven: it relies on no external forcing or fine tuning of any
parameters.  To our knowledge, this is the first self-consistent model
for the spontaneous onset of fast reconnection.

To see why transitions to fast reconnection are generic to the
reconnection process, consider a macroscopic current sheet with a
small but non-negligible resistivity.  Without small scale structure,
the Hall term in Ohm's law is unimportant, so the system undergoes
Sweet-Parker reconnection.  Since the resistivity is small, the
dissipation region is embedded within the macroscopic current sheet.
The width $\delta$ of the Sweet-Parker current layer is given by
\citep{Sweet58,Parker57}
\begin{equation}
\frac{\delta}{L} \sim \left(\frac{\eta c^{2}}{4 \pi c_{Aup}
  L}\right)^{1/2} \label{spd}
\end{equation}
where $c_{Aup} = B_{up} / (4 \pi m_{i} n)^{1/2}$ is the Alfv\'en speed
evaluated using the magnetic field $B_{up}$ just upstream of the
dissipation region and $L$ is the macroscopic length of the
Sweet-Parker current sheet.  During reconnection, stronger magnetic
field lines are convected into the dissipation region by the inflow,
causing a decrease in the Sweet-Parker layer width.  When the
dissipation region becomes thinner than $d_{i}$ a transition ensues.

In the following Section, we review the salient properties of
Sweet-Parker and Hall reconnection.  In \S~\ref{sec-sims}, we
describe the numerical simulations and their results, and we discuss
implications for the onset of solar flares in \S~\ref{sec-disc}.

\section{SWEET-PARKER AND HALL RECONNECTION}
\label{sec-spandhall}

In the Sweet-Parker model \citep{Sweet58,Parker57}, a steady state is
reached when the convective inflow of magnetic field lines is balanced
by diffusion of field lines towards the X-line,
\begin{equation}
\frac{v_{in}}{\delta} \sim \frac{\eta c^{2}}{4 \pi \delta^{2}},
\label{SPresult}
\end{equation}
where $v_{in}$ is the inflow speed.  From continuity, $v_{in} L \sim
v_{out} \delta$, where $v_{out}$ is the outflow speed.  Therefore, the
inflow Alfv\'en Mach number $M_{A} = v_{in} / c_{Aup}$ (a measure of
the reconnection rate) is given by the right hand side of equation
(\ref{spd}) since $v_{out} \sim c_{Aup}$.  The extreme elongation of
the dissipation region due to $\eta$ being very small for most plasmas
of interest throttles Sweet-Parker reconnection.  For solar flares,
the Sweet-Parker prediction of $M_{A} \sim 2 \times 10^{-7}$ is far
smaller than inferred from observations, where we used typical
parameters of $n \sim 10^{9}\;{\rm cm}^{-3}$ for the preflare density,
$B_{0} \sim 100 \;{\rm G}$ for the preflare coronal magnetic field, $L
\sim 10^{4}\;{\rm km}$ for a typical flux tube length, and a classical
resistivity of $\eta \sim 2 \times 10^{-16}\;{\rm sec}$ based on a
temperature of $T \sim 10^{6}\;{\rm K}$ \citep{Priest02}.

The physics of Hall reconnection is fundamentally different from that
of Sweet-Parker.  The motion of ions decouples from that of the
electrons and the magnetic field at a distance $d_{i}$ from the
X-line.  The electrons remain frozen-in to the magnetic field down to
the electron skin depth $d_{e} = c / \omega_{pe}$.  Where the species
are decoupled, the Hall term in Ohm's law introduces whistler and/or
kinetic Alfv\'en waves (depending on the plasma $\beta$) into the
system \citep{Mandt94,Rogers01}.  Both waves are dispersive with
$\omega \propto k^{2}$.  The dispersive property of these waves causes
the outflow jet from the X-line to open as discussed by Petschek
\citep{Rogers01}.  In the absence of dispersive waves, reconnection is
slow as in the Sweet-Parker model.

Numerical simulations \citep{Shay99,Huba04,Shay04} have shown that the
inflow speed for steady-state Hall reconnection is
\begin{equation}
v_{in} \sim 0.1 c_{Aup} \label{hallresult}
\end{equation}
({\it i.e.}, $M_{A} \sim 0.1$).  This result has been found to be
independent of electron mass \citep{Shay98b,Hesse99}, system size
\citep{Shay99}, and dissipation mechanism \citep{Birn01}.  Thus, we
expect a dramatic increase in the reconnection rate when a transition 
from Sweet-Parker to Hall reconnection occurs.

\section{NUMERICAL SIMULATIONS AND RESULTS}
\label{sec-sims}

We perform numerical simulations using the massively parallel
compressible two-fluid code {\sc F3D} \citep{Shay04} in a periodic
two-dimensional domain.  The initial equilibrium magnetic field is a
double current sheet configuration given by one period of a sine
sheet, $B_{x0}(y) = B_{0} \cos(2 \pi y/ L_{y})$, where $L_{y}$ is the
size of the domain in the inflow direction, with pressure balance
enforced by a non-uniform density profile, $n(y) = n_{0} + (B_{0}^{2}
/ 8 \pi T_{0}) \sin^{2}(2 \pi y/ L_{y})$.  Here, $n_{0}$ is a constant
corresponding to the density at the edge of the domain and $T_{0} =
B_{0}^{2} / 4 \pi n_{0}$ is the temperature, assumed constant and
uniform for simplicity.  The initial density at the center of the
current sheet is, therefore, $1.5 n_{0}$.  We impose no initial guide
field.  Lengths are normalized to the ion skin depth $d_{i0}$ based on
the density $n_{0}$ at the edge of the computational domain, not the
center of the X-line, which we denote as $d_{iX}$.  Magnetic field
strengths, velocities, times, and resistivities are normalized to
$B_{0}$, the Alfv\'en speed $c_{A0}$ based on $B_{0}$ and $n_{0}$, the
ion cyclotron time $\Omega_{ci}^{-1} = (e B_{0} / m_{i} c)^{-1}$, and
$\eta_{0} = 4 \pi c_{A0} d_{i0} / c^{2}$, respectively.

The computational domain is of size $L_{x} \times L_{y} = 409.6 d_{i0}
\times 204.8 d_{i0}$ with a cell size of $0.1 d_{i0} \times 0.1
d_{i0}$.  There is no viscosity, but fourth order diffusion with
coefficient $2 \times 10^{-5}$ is used in all of the equations to damp
noise at the grid scale.  An electron mass of $m_{e} = m_{i} / 25$ is
used.  Although this value is unrealistic, the electron mass only
controls dissipation at the electron scales which does not impact the
rate of Hall reconnection.  A small coherent perturbation ${\bf B}_{1}
= - (0.004 B_{0} L_{y} / 2 \pi) {\bf \hat{z}} \times \nabla [\sin(2 
\pi x / L_{x}) \sin^{2}(2 \pi y / L_{y})]$ is used to initiate 
reconnection.  The resistivity is taken to be uniform.  Simulations 
are performed with $\eta = 0.0025 \eta_{0}$ and $0.0090 \eta_{0}$, 
both of which exhibit transitions to fast reconnection.  We present 
results from the $\eta = 0.0025 \eta_{0}$ simulation, which was 
initialized from the $\eta = 0.0090 \eta_{0}$ simulation at $t = 
5.364 \;{\rm k}\Omega_{ci}^{-1}$.  Initializing the simulation in this
way introduces transient behavior, but it dies away (by $t \sim 11
\;{\rm k}\Omega_{ci}^{-1}$) before small scale dynamics become 
important.

When the system is evolved in time, the Hall effect is initially very
small because the width of the current layer $L_{y} / 2 = 102.4
d_{i0}$ is large compared to $d_{iX}$, so the system evolves
essentially as it would in pure resistive MHD.  A Sweet-Parker current
layer develops, as we will demonstrate later.  The ion and electron
inflow velocities, measured as the maximum value of the inflow into
the X-line for each species, are plotted as a function of time late in
the simulation in Figure~\ref{spontplots}a.  Up until $t \sim 18
\;{\rm k}\Omega_{ci}^{-1}$, the electrons and ions are coupled as
expected in MHD.  The inflow speed is very small, but is slowly rising
due to a gradual increase in the upstream magnetic field strength
$B_{up}$ as stronger magnetic fields are convected into the
dissipation region.  Figure~\ref{spontplots}b shows the slow increase
in $B_{up}$, measured just upstream of the current layer in the
simulation.

When the ions decouple from the electrons, the inflow speeds begin to
increase dramatically and the system begins a transition to Hall
reconnection.  This transition initiates when the width of the current
layer $\delta$ falls below $d_{iX}$, as is shown in
Figure~\ref{spontplots}c.  The thick solid line is $d_{iX}$ as a
function of time.  After decoupling, one must distinguish between the
electron and ion current sheet widths, which we denote as $\delta_{e}$
and $\delta_{i}$, respectively.  The solid line is $\delta_{e}$,
determined by the half width at half maximum of the total current
layer.  The dashed line is $\delta_{i}$, determined by the greater of
$\delta_{e}$ and the half width at half maximum of the total inflow
current.  The latter becomes non-zero where the electrons and ions
decouple, and is therefore a measure of the edge of the ion
dissipation region.  One can see $\delta_{i}$ decreasing from large
scales (larger than $d_{iX}$) as the upstream magnetic field
increases, and the transition begins when it is of the order of
$d_{iX}$.

\begin{figure}
\plotone{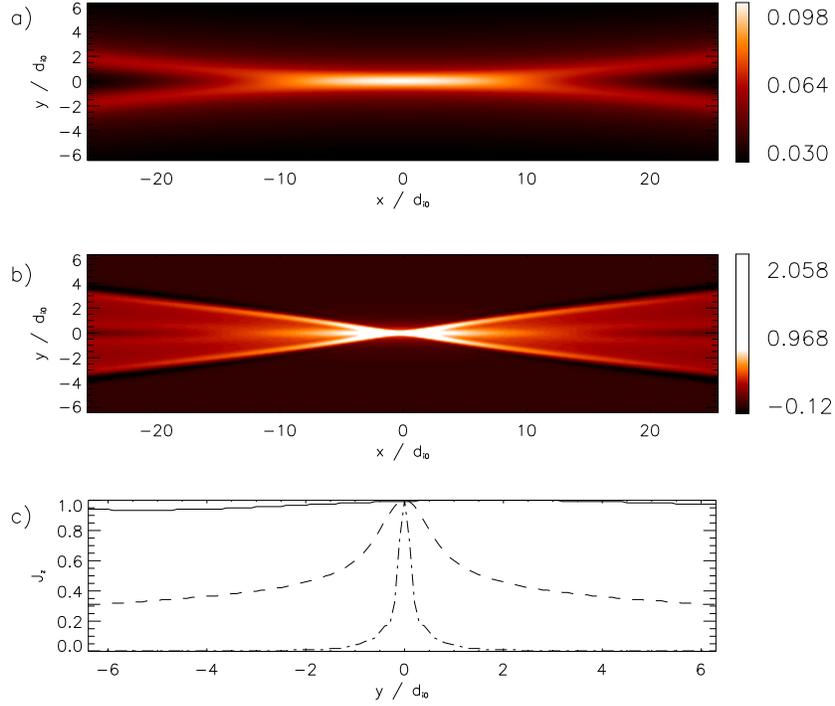}
\caption{\label{cursheet} (Color online) Grayscale plot of the current
sheet during (a) Sweet-Parker reconnection (at $t = 11.4 \;{\rm k} 
\Omega_{ci}^{-1}$) and (b) Hall reconnection (at $t = 19.6 \;{\rm k} 
\Omega_{ci}^{-1}$).  (c) Cuts across the X-line for the same two 
sheets (dashed and dot-dashed, respectively) normalized to its maximum 
value.  The initial current sheet profile is the solid line.  Notice 
the color table for b) has been skewed for greater contrast and the 
amplitude of the current density is vastly different for the two 
sheets.}
\end{figure}

Finally, to verify that the system is undergoing Sweet-Parker before
the transition, and Hall reconnection after, we must check the
validity of the inflow speed predictions from equations
(\ref{SPresult}) and (\ref{hallresult}).  The thick solid line of
Figure~\ref{spontplots}d shows $v_{in}$ as a function of time.  The
dashed line is the Sweet-Parker prediction from equation
(\ref{SPresult}) ($v_{in} \sim \eta / \delta$ in code units), while
the thin solid line is the Hall reconnection prediction with a
constant coefficient of 0.17, which is of the order of $\sim 0.10$ 
as expected from equation (\ref{hallresult}).  Clearly, up until 
about $t \sim 18 \;{\rm k}\Omega_{ci}^{-1}$, there is excellent 
agreement with the Sweet-Parker result.  A grayscale plot of the 
current layer during the Sweet-Parker phase (at $t = 11.4 \;{\rm k} 
\Omega_{ci}^{-1}$) is shown in Figure~\ref{cursheet}a, showing the 
characteristic elongated dissipation region (similar to those 
observed with pure MHD simulations by \citet{Jemella04}).  After a 
relatively brief transition time lasting until $t \sim 19.5\;{\rm k} 
\Omega_{ci}^{-1}$, the inflow speed is well modeled by the Hall 
prediction.  A grayscale plot of the current layer during the Hall 
phase (at $t = 19.6 \;{\rm k} \Omega_{ci}^{-1}$) is shown in 
Figure~\ref{cursheet}b, showing the open outflow configuration 
characteristic of Hall reconnection.  We observe a large enhancement 
of the quadrupolar structure in the out of plane magnetic field, a 
signature of Hall reconnection \citep{Mandt94}.  Cuts across the 
current sheet at the X-line normalized to its maximum value are 
plotted as the dashed line and dot-dashed lines in 
Figure~\ref{cursheet}c, showing that $\delta_{e}$ falls to $d_{e} =
0.2 d_{i}$ during Hall reconnection, as is expected when electron
inertia provides the dissipation.  For comparison, the solid line is a
cut across the initial equilibrium current sheet.

\section{DISCUSSION}
\label{sec-disc}

The spontaneous onset model presented here provides a possible
explanation of why reconnection sites in weakly collisional plasmas
are apparently quiet for a long time as magnetic energy accumulates
before a sudden onset of fast magnetic reconnection releases it.  A
rigorous comparison of this model with flare observations is
challenging because the dissipation regions associated with the
transition to fast reconnection are much narrower than can be resolved
with satellite or ground-based observations.  We can, however, compare
some basic predictions with observations.

First, are macroscopic current sheets in the corona wide compared to
the ion skin depth?  Using values of the plasma parameters in a solar
flare from \S~\ref{sec-spandhall} gives an ion skin depth of only
$d_{i} \sim 7 \times 10^{2}\;{\rm cm}$, far narrower than expected
macroscopic current sheets in the corona.  At present, current sheets
in the corona are inaccessible to observations, though reasonable
scales for the current sheet width $W_{s}$ may be 100-1,000 km.

Second, taking $\eta$ as a given, what is the critical upstream
magnetic field strength $B_{*}$ which would make the Sweet-Parker
current layer width equal to $d_{i}$?  Setting $\delta = d_{i}$ in
equation (\ref{spd}), we find
\begin{equation}
B_{*} \sim \sqrt{4 \pi m_{i} n_{0}} \left(\frac{\eta c^{2}}{4 \pi
d_{i}^{2}} L\right) \sim 5\;{\rm G}
\end{equation}
using the values from \S~\ref{sec-spandhall}.  This is accessible
during reconnection in the corona.

Third, what is the time scale for the quiet time $\tau_{q}$, during
which Sweet-Parker reconnection could be active but magnetic energy
could accumulate?  Since the field is frozen-in outside of the
dissipation region, it is the time it takes for a field of strength
$B_{*}$ to be convected in by the inflow,
\begin{equation}
\tau_{q} = \int \frac{d \xi}{v_{in}},
\end{equation}
where $\xi$ is the distance upstream from the X-line.  This can be
approximated using $v_{in} \sim (\eta c^{2} / 4 \pi c_{A} L)^{1/2}$
and by assuming a linear profile in the magnetic field $B = B_{0} \xi
/ W_{s}$ in $c_{A}$.  Integrating from $\xi \sim B_{*} W_{s} / B_{0}$
to approximately zero gives
\begin{equation}
\tau_{q} \sim 2 W_{s} \sqrt{\frac{4 \pi L}{\eta c^{2} c_{A0}}
\frac{B_{*}}{B_{0}}} \sim \left( \frac{W_{s}}{100\;{\rm km}}
\right) 4 \times 10^{4}\;{\rm sec},  \label{qtime}
\end{equation}
where $c_{A0}$ is the Alfv\'en speed based on $B_{0}$.  The numerical
factor is about 11 hours, which is a reasonable time scale for the
accumulation of magnetic energy due to footpoint motion in the 
photosphere \citep{Dahlburg05}.

\begin{figure}
\epsscale{.70}
\plotone{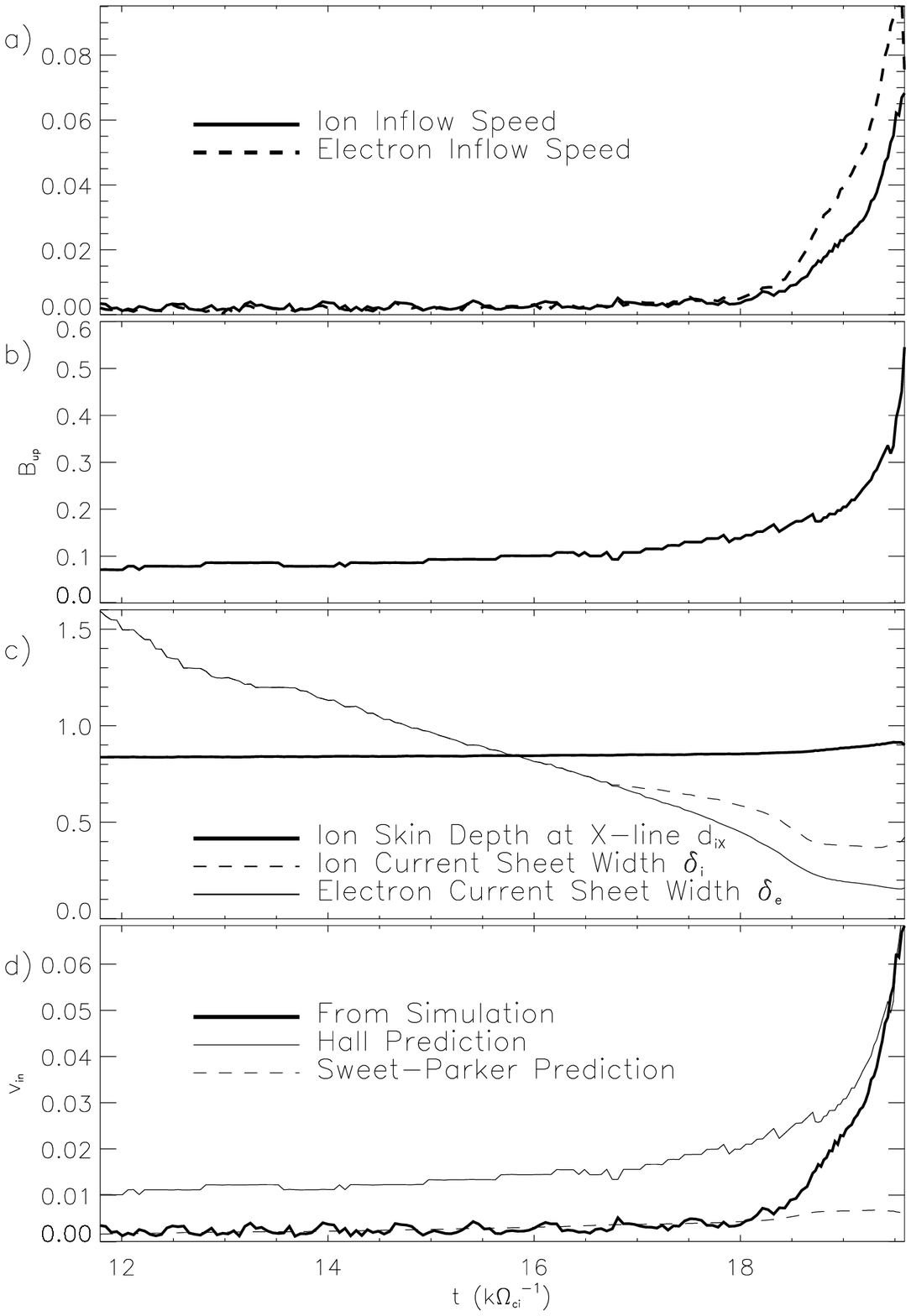}
\caption{\label{spontplots} Time dependence (in thousands of ion
cyclotron times) of the (a) ion (solid) and electron (dashed) inflow
velocities, (b) upstream magnetic field strength $B_{up}$, (c)
electron (thin solid) and ion (dashed) current sheet widths
$\delta_{e}$ and $\delta_{i}$ and ion skin depth (thick solid) at the
X-line $d_{iX}$, and (d) ion inflow velocity from the simulation
(thick solid), with Sweet-Parker theory (dashed, from
eq.~[\ref{SPresult}]) and Hall theory (thin solid, from
eq.~[\ref{hallresult}] with 0.17 replacing 0.10).}
\end{figure}

The time it takes for the transition from Sweet-Parker to Hall
reconnection, corresponding to the time from onset until maximum flare
signal, can be bounded above by the convective time across the
Sweet-Parker current sheet $\delta / v_{in}$, which at the transition
time is the same as the resistive time across the layer $(\eta c^{2} /
4 \pi \delta^{2})^{-1}$ and the convective time along the layer $L /
v_{out}$.  For our simulation, the resistive time is $\sim 400
\Omega_{ci}^{-1}$, which compares reasonably well with the observed
time of the transition (see Figure~\ref{spontplots}d).  For solar
flare parameters, the resistive time across the layer is approximately
$28 \;{\rm sec}$, which is comparable to the onset times seen in
flares \citep{Priest02}.  The predicted observable parameters are
quite consistent with solar flare phenomena.

The present simulations do not include the effect of an out of plane
(guide) field, the more generic configuration for magnetic
reconnection.  It was conjectured \citep{Cassak05} that the transition
to fast reconnection in the presence of a guide field is also
catastrophic, but occurs when the width of the current layer reaches
the ion Larmor radius $\rho_{s} = c_{s} / \Omega_{ci}$, where $c_{s}$
is the ion sound speed, instead of the ion skin depth $d_{i}$.  This
is because $\rho_{s}$ is the scale where dispersive (kinetic Alfv\'en)
waves become important in the presence of a guide field
\citep{Rogers01}.  Interestingly, recent laboratory experiments at the
Versatile Toroidal Facility \citep{Egedal00} have observed spontaneous
reconnection, and preliminary diagnostics suggest that the width of
the current layer at onset is very close to their value of the ion
Larmor radius $\rho_{s}$ (Egedal, private communication).

Finally, \citet{Longcope05} recently observed an active region
reconnect with a nearby flux loop as it emerged from the corona.  A
phase of slow reconnection was observed for $\sim 24\;{\rm hr}$,
during which magnetic energy accumulated in the corona.  This was
followed by fast reconnection lasting $\sim 3\;{\rm hr}$.  The onset
was sudden with no visible trigger mechanism observed.  The energy
released during fast reconnection was shown to be comparable to the
energy accumulated during slow reconnection.  Based on parameters
inferred from the observations (a loop voltage of $10^{9} \;{\rm V}$,
a current sheet depth of $2 \times 10^{5} \;{\rm km}$, a sheet length
of $L \sim 3 \times 10^{4} \;{\rm km}$, a sheet current of $I \sim
1.34 \times 10^{11} \;{\rm A}$, and a density of $n \sim 10^{9} \;
{\rm cm^{-3}}$), the fast reconnection rate was $M_{A} \sim 0.05$,
based on a reconnection electric field of $E \sim 5 \;{\rm V/m}$ and a
reconnecting magnetic field of $B_{0} \sim 4 \;{\rm G}$, consistent
with Hall reconnection.  These observations provide solid evidence for
the accumulation of magnetic energy during a slow reconnection phase
followed by a spontaneous onset of fast reconnection, as proposed
here.

\acknowledgments

This work has been supported by NSF Grant No.~PHY-0316197 and DOE
Grant Nos.~ER54197 and ER54784.  Computations were carried out at the
National Energy Research Scientific Computing Center.


\clearpage

\clearpage

\end{document}